\documentclass[aps,prb,twocolumn,amsmath,amssymb,superscriptaddress,showpacs]{revtex4}
\usepackage{graphicx}
\usepackage{color}
\usepackage{Jonasmacros}
\usepackage{bm}

\begin{document}
\date{\today}
\title{Theory of spin-polarized scanning tunneling microscopy applied to local spins}
\author{J. Fransson}
\email{Jonas.Fransson@fysik.uu.se}
\affiliation{Department of Physics and Materials Science, Uppsala University, Box 530, SE-751 21\ \ Uppsala}
\author{O. Eriksson}
\affiliation{Department of Physics and Materials Science, Uppsala University, Box 530, SE-751 21\ \ Uppsala}
\author{A. V. Balatsky}
\email{avb@lanl.gov}
\affiliation{Theoretical Division, Los Alamos National Laboratory, Los Alamos, New Mexico 87545, USA}
\affiliation{Center for Integrated Nanotechnology, Los Alamos National Laboratory, Los Alamos, New Mexico 87545, USA}

\begin{abstract}
We provide a theory for scanning tunneling microscopy and spectroscopy using a spin-polarized tip. It it shown that the tunneling conductance can be partitioned into three separate contributions, a background conductance which is independent of the local spin, a dynamical conductance which is proportional to the local spin moment, and a conductance which is proportional to the noise spectrum of the local spin interactions. The presented theory is applicable to setups with magnetic tip and substrate in non-collinear arrangement, as well as for non-magnetic situations. The partitioning of the tunneling current suggests a possibility to extract the total spin moment of the local spin from the dynamical conductance. The dynamical conductance suggests a possibility to generate very high frequency spin-dependent ac currents and/or voltages. We also propose a measurement of the dynamical conductance that can be used to determine the character of the effective exchange interaction between individual spins in clusters. The third contribution to the tunneling current is associated with the spin-spin correlations induced by the exchange interaction between the local spin moment and the tunneling electrons. We demonstrate how this term can be used in the analysis of spin excitations recorded in conductance measurements. Finally, we propose to use spin-polarized scanning tunneling microscopy for detailed studies of the spin excitation spectrum.
\end{abstract}
\maketitle

\section{Introduction}
\label{sec-intro}
Pushing the limits for detection of electronic, magnetic, and vibrational properties towards the quantum limit requires appropriate experimental tools and techniques. In particular, single atomic spins \cite{gambardella2003,rugar2004,franssonNT2008} and magnetic nanostructures consisting of few magnetic atoms \cite{heinrich2004,hirjibehedin2006,hirjibehedin2007,kitchen2007,balashov2009} on nonmagnetic substrates are frequently studied as model systems for miniature data storage devices, spintronics applications, and qubits which are crucial for quantum information technology. Being well-defined and controllable on the atomic scale, they ideally serve their purpose for studies of fundamentals of their local properties and interactions.

The scanning tunneling microscope (STM) was invented in the 1980's \cite{binnig1981} for the purpose of imaging metallic surfaces with atomic resolution, and it was theoretically described by Tersoff and Hamann,\cite{tersoff1983} relating the tunneling conductance to the density of states (DOS) of the local environment. More than being a scanning tool, one can park the tip over an object and perform differential conductance measurements in order to reveal information of the local electronic structure. Using the STM equipment as a means for spectroscopical measurements was previously discussed, see e.g. Refs. \onlinecite{selloni1985,lang1986,ukraintsev1996,koslowski2007,passoni2007,passoni2009}, the approach that is commonly known as scanning tunneling spectroscopy (STS). This technique has since been successfully applied in several directions, e.g. detection of noise,\cite{manassen1989} indirect measurements of the Kondo effect,\cite{wahl2007} and the observation of exchange splitting.\cite{lee2004}

Extensions of the STM/STS techniques have been provided by using a spin-polarized tip (SP-STM/SP-STS), e.g. magnetic CrO$_2$ tips,\cite{wiesendanger1990} or e.g nonmagnetic W tips which are coated with a ferromagnetic metal, e.g. Fe,\cite{heinze2000} or an anti-ferromagnetic metal, e.g. Cr.\cite{kubetzka2002} To this end, the theory by Tersoff and Hamann was extended to also account for the spin-polarization in the tip and substrate.\cite{reittu1997,wortmann2001,heinze2006} The latter formulation was for instance used in Ref. \onlinecite{meier2008} in analyzing experimental SP-STM results of spin-moments for single Co atoms on a Pt substrate. The lack of explicit reference to the spin moments of the adatoms located on the substrate in this theory, however, calls for an advancement in the theoretical formulation. In this paper we present a theory which suggest to use the SP-STM technique for directly extracting quantitative information about the magnetic moments of nano-scale objects.

The theory presented here is discussed in the context of recent experimental observations, using both STM and STS approaches and using both a nonmagnetic and spin-polarized tip. Our discussion will be cast in the light of the theoretical description of the SP-STM, particularly for measurements performed in presence of local spin moments, $\bfS_n$, located at $\bfr_n$ on the substrate. We find that indeed STM tunneling is in principle capable of detecting single spin and moreover detect the spin orientation. This conclusion can be drawn, for instance, when we consider the tunneling electrons to be interacting with the local spins through exchange. For this type of interaction mechanism, the tunneling matrix element can be separated into one spin-independent and one spin-dependent component. Under such conditions, we find that the tunneling conductance can be separated into three components, of which the first provides a conductance depending on the electron and magnetic densities of the tip and substrate, the second yields a conductance which is directly proportional to the local spin moment, and the third being proportional to the noise produced by the local spin fluctuations. The last contribution to the current, or conductance, was recently discussed in Refs. \onlinecite{franssonNL2009,rossier2009,persson2009}.

We point out that we are interested in the qualitative effects caused by the presence of local spin, or magnetic, moments located on the substrate surface. For this reason, we will build upon previous results concerning the matrix elements for the tunneling electrons between the tip and substrate. In particular, the spin-dependence in the tunneling current generated by the local spin moment via exchange interaction is of interest, while other features that are pertinent to the tunneling, e.g. geometry of the tip, surface, and adsorbate, bias voltage dependence of the matrix element, etc, will be treated with less accuracy.

We begin the paper by a derivation of the tunneling current and the corresponding (differential) conductance in Sec. \ref{sec-derivation}. We continue by discussing the main properties of the conductance contributions in Sec. \ref{sec-analysis}, and thereafter we discuss possible experimental measurements suggested by the theoretical results in Sec. \ref{sec-exp}, and we conclude the paper in Sec. \ref{sec-conclusion}.

\section{Probing the local spin moment}
\label{sec-derivation}
The wave function of the tunneling electrons in the tip, separated from the substrate by distance $d$, has an exponentially small overlap with the substrate electron wave functions. The spin-dependent tunneling matrix element can be calculated using Bardeen's result,\cite{bardeen1967} and is given by \cite{balatsky2002,nussinov2003,zhu2003}
\begin{align}
\Gamma_n=&
	\Gamma_0 \exp{\left(-\sqrt{\frac{\Phi-J\bfS_n(t)\cdot\bfsigma}{\Phi_0}}\right)},
\label{eq-me}
\end{align}
which is understood as a matrix in spin space. Here, $\bfsigma=(\sigma^x,\sigma^y,\sigma^z)$ is the Pauli matrix vector, whereas $\Gamma_0$ describes the spin-independent tunneling in absence of $J$. $\Phi$ is the tunneling barrier height, while $\Phi_0=\hbar^2/8md^2$ is the energy related to the distance $d$ between the tip and surface. In principle, the spin-dependent tunneling matrix element depends on energy, position, applied bias voltage, and quantum numbers of the electron states, which have been discussed extensively in the literature.\cite{selloni1985,lang1986,hormandinger1994,ukraintsev1996,tekman1992,bracher1997,wortmann2005} In the present study we shall omit those dependencies, however, for the sake of focusing on the dependence of the tunneling conductance on the local spin moments.

The exchange energy $J|\bfS|$ is small compared to the barrier height, so that we can expand the exponent and find the effective tunneling matrix element
\begin{align}
\bfT_n=&
	T_0+T_1\bfsigma\cdot\bfS_n,
\end{align}
where
\begin{subequations}
\begin{align}
T_0=&\Gamma_0e^{-\sqrt{\Phi/\Phi_0}}\cosh\frac{J|\bfS|}{2\Phi}\sqrt{\frac{\Phi}{\Phi_0}},
\\
T_1=&\Gamma_0e^{-\sqrt{\Phi/\Phi_0}}\sinh\frac{J|\bfS|}{2\Phi}\sqrt{\frac{\Phi}{\Phi_0}},
\end{align}
\end{subequations}
such that $T_1/T_0\sim J/\Phi$. For metals and semiconductors it is reasonable to use $J\sim0.1$ eV,\cite{bhattacharjee1992} while the tunneling barrier $\Phi\sim1$ eV, giving typical values of $T_1/T_0\sim1/10$. In the following discussion, the tunneling rates $T_0$ and $T_1$ are treated as constants, except that we allow $T_1$ to carry a spatial dependence which is related to the positions of the local spin moments. More details of this description of the tunneling matrix elements can be found in e.g. Refs. \onlinecite{balatsky2002,nussinov2003,zhu2003}.

We next assume that the substrate surface is metallic for which $\Hamil_{\sub}=\sum_{\bfk\sigma}\leade{\bfk}\cdagger{\bfk}\cc{\bfk}$ is sufficient, where $\cdagger{\bfk}$ creates a surface electron with energy $\leade{\bfk}$, momentum $\bfk$, and spin $\sigma$. The energy-momentum dispersion relation need not be of free-electron character but may assume any general form, for which the specific details are unimportant for the present derivation. The energies $\leade{\bfk}$ are given relative to the Fermi level $\dote{F}$, which is common for the system as a whole. We associate the electronic and magnetic densities $N(\bfr,\dote{})$ and $\bfM(\bfr,\dote{})$, respectively, with the electrons in the substrate. For simplicity, we will assume in the following that those densities are slowly varying with energy. Analogously, we model the electrons in the tip by $\Hamil_{\tip}=\sum_{\bfp\sigma}\leade{\bfp}\cdagger{\bfp}\cc{\bfp}$, and define its corresponding electronic and magnetic densities $n(\dote{})$ and $\bfm(\dote{})$, respectively.

In general, the magnetic moments of the substrate and tip may be in a non-collinear arrangement. Thus, defining the $z$-direction of the global reference frame in e.g. the spin-quantization axis of the tip, the operators of the substrate are transformed according to
\begin{align}
\left(\begin{array}{c}
	\cs{\bfk\up} \\ \cs{\bfk\down}
\end{array}\right)
=&
	\left(\begin{array}{cc}
		\cos(\theta/2)e^{-i\phi/2} & -\sin(\theta/2)e^{-i\phi/2} \\
		\sin(\theta/2)e^{i\phi/2} & \cos(\theta/2)e^{i\phi/2} \\
	\end{array}\right)
	\left(\begin{array}{c}
		\cs{\bfk+} \\ \cs{\bfk-}
	\end{array}\right)
\end{align}
where $\phi$ and $\theta$ are the azimuthal and polar angles between the local substrate reference frame (spins $s=\pm$) and the global one (spins $\sigma=\up,\down$).

Tunneling of electrons between the tip and substrate in the presence of the local spin moments is captured by the model
\begin{align}
\Hamil_T=&	
	\sum_{\stackrel{\scriptstyle \bfp\bfk n}{\sigma\sigma'}}
		\cdagger{\bfp}
			[T_0\delta_{\sigma\sigma'}
			+T_1(\bfr-\bfr_n)\bfsigma_{\sigma\sigma'}\cdot\bfS_n]
\nonumber\\&\times
		\cs{\bfk\sigma'}
		e^{i\bfk\cdot\bfr+ieVt}
		+H.c.,
\end{align}
where the bias voltages applied across the junction is denoted by $eV$. Due to the local nature of the spins, we have included a spatial dependence in the interacting tunneling rate, and we use e.g. $T_1(\bfr-\bfr_n)=T'\exp{(-|\bfr-\bfr_n|/\lambda)}$, where $\lambda$ is the decay length.

The charge current running between the tip and the substrate is derived using non-equilibrium Green functions on the Keldysh contour, starting from the fundamental relation $I(t)=-e\dt\sum_{\bfp\sigma}\av{n_{\bfp\sigma}}$, where $n_{\bfp\sigma}=\cdagger{\bfp}\cc{\bfp}$, giving
\begin{align}
I(\bfr,t,V)=&
	-\frac{2e}{\hbar}\im\sum_{\stackrel{\scriptstyle \bfp\bfk n}{\sigma\sigma'}}
		\av{\cdagger{\bfp}(t)\hat{T}_{\sigma\sigma'}(\bfr_n,t)\cs{\bfk\sigma'}(t)}
\nonumber\\&\times
		e^{i\bfk\cdot\bfr+ieVt},
\end{align}
where we have defined $\hat{T}_{\sigma\sigma'}(\bfr_n,t)=T_0\delta_{\sigma\sigma'}+T_1(\bfr-\bfr_n)\bfsigma_{\sigma\sigma'}\cdot\bfS_n(t)$. The current expresses the rate of change of the expectation value of $n_{\bfp\sigma}$, corresponding to the number of charges, $e$, in the tip. Expanding the average in the above expression according to $\av{A(t)}\approx(-i)\int_C\av{\com{A(t)}{\Hamil_T(t')}}dt'$, and by converting to real times, the current can be written as
\begin{align}
I(\bfr,t,V)=&
	\frac{2e}{\hbar}\re
		\sum_{\stackrel{\scriptstyle \bfp\bfp'}{\bfk\bfk'}}
		\sum_{\stackrel{\scriptstyle \sigma\sigma'}{\sigma''\sigma'''}}
		\sum_{nm}
		\int_{-\infty}^t
		e^{i(\bfk-\bfk')\cdot\bfr+ieV(t-t')}
\nonumber\\&\times
		\langle[\cdagger{\bfp}(t)\hat{T}_{\sigma\sigma'}(\bfr_n,t)\cs{\bfk\sigma'}(t),
\nonumber\\&
			\csdagger{\bfk'\sigma'''}(t')\hat{T}_{\sigma'''\sigma''}(\bfr_m,t')\cs{\bfp'\sigma''}(t')]
		\rangle
		dt'.
\label{eq-I}
\end{align}
Here, notice the appearance of the commutator taken between the operators inside the average. Separating the tunneling operator $\hat{T}$ into its components $T_0$ and $T_1$, we find that the current can be naturally written as a sum of three terms, i.e. $I(t)=\sum_{i=0}^2I_i(t)$. We consider the differential conductance $\partial I(\bfr,t,V)/\partial V=\sum_{i=0}^2\partial I_i(\bfr,t,V)/\partial V$ of stationary source drain voltages since such conditions are predominant in experimental situations.

\subsection{Background conductance}
The first contribution from the average in Eq. (\ref{eq-I}) is given by
\begin{align}
T_0^2&
	\delta_{\sigma\sigma'}\delta_{\sigma'''\sigma''}
	\av{\com{\cdagger{\bfp}(t)\cc{\bfk}(t)}{\cdagger{\bfk'}(t')\cc{\bfp'}(t')}}
\nonumber\\=&
T_0^2\delta_{\sigma\sigma'}\delta_{\sigma'''\sigma''}
	(
		\av{\cdagger{\bfp}(t)\cc{\bfp'}(t')}\av{\cc{\bfk}(t)\cdagger{\bfk'}(t')}
\nonumber\\&
		-\av{\cc{\bfp'}(t')\cdagger{\bfp}(t)}\av{\cdagger{\bfk'}(t')\cc{\bfk}(t)}
	),
\end{align}
where we have decoupled the averages in the integral in terms of correlation functions for the substrate and tip electrons. Under the assumption that scattering between different momentum and spin states can be neglected, and introducing the notations $g_{\bfkappa\sigma}^<(t,t')=i\av{\cdagger{\bfkappa}(t')\cc{\bfkappa}(t)}$ and $g_{\bfkappa\sigma}^>(t,t')=(-i)\av{\cc{\bfkappa}(t)\cdagger{\bfkappa}(t')}$, the first contribution to the tunneling current is defined by
\begin{align}
I_0(\bfr,t,V)=&
	\frac{2eT_0^2}{\hbar}\re\sum_{\bfp\bfk}\sum_\sigma
		\int_{-\infty}^t
		[
			g^<_{\bfp\sigma}(t',t)
			g^>_{\bfk\sigma}(t,t')
\nonumber\\&
			-g^>_{\bfp\sigma}(t',t)
			g^<_{\bfk\sigma}(t,t')
		]
		e^{ieV(t-t')}
		dt'.
\end{align}
Here, the lesser and greater correlation functions of the tip electrons are given by 
\begin{align}
g^{</>}_{\bfp\sigma}(t,t')=&
	(\pm i)f(\pm\leade{\bfp})e^{-i\leade{\bfp}(t-t')},
\end{align}
respectively, and analogously for the correlation functions of the substrate electrons. Here, $f(x)$ is the Fermi function. Thus, performing the time-integration and replacing the momentum summations by integrations over the corresponding spin-resolved density of electron states (DOS) $n_\sigma(\dote{})$ and $N_\sigma(\bfr,\dote{})$ in the tip and substrate, respectively, we finally arrive at  the tunneling current
\begin{align}
I_0(\bfr,t,V)=&
	\frac{2\pi eT_0^2}{\hbar}\sum_\sigma\int
		n_\sigma(\dote{}-eV)N_\sigma(\bfr,\dote{})		
\nonumber\\&\times
		[f(\dote{}-eV)-f(\dote{})]
		d\dote{}
\nonumber\\=&
	\frac{2\pi^2 eT_0^2}{h}\int
		[f(\dote{})-f(\dote{}-eV)]
		[n(\dote{}-eV)
\nonumber\\&\times
		N(\bfr,\dote{})
		+\bfm(\dote{}-eV)\cdot\bfM(\bfr,\dote{}))
		d\dote{}.
\end{align}
The last expression is obtained by noting that $n_\sigma=(n+\sigma m_z)/2$ and $N_\sigma=(N+\sigma M_z\cos\theta)/2$ (the factor $\sigma=\pm1$) which thus provides a description that is independent of the coordinate system.

We are, ultimately, interested in the tunneling (differential) conductance, $\partial I(\bfr,V)/\partial V$, to which $I_0$ contributes
\begin{align}
\frac{\partial I_0(\bfr,t,V)}{\partial V}
	=&
	\frac{\pi^2\sigma_0T_0^2}{4k_BT}
	\int
		\cosh^{-2}\frac{\dote{}-eV}{2k_BT}
		[n(\dote{}-eV)
\nonumber\\&\times\vphantom{\int}
		N(\bfr,\dote{})
		+\bfm(\dote{}-eV)\cdot\bfM(\bfr,\dote{})]
		d\dote{}
\nonumber\\
\rightarrow&\vphantom{\int}
	\pi^2\sigma_0T_0^2
		[n(\dote{F}-eV)
		N(\bfr,\dote{F})
\nonumber\\&\vphantom{\int}
		+\bfm(\dote{F}-eV)\cdot\bfM(\bfr,\dote{F}))
		,\
		T\rightarrow0,
\label{eq-dI0}
\end{align}
where $\sigma_0=2e^2/h$ is the fundamental conductance unit, whereas $k_B$ and $T$ is the Boltzmann constant and temperature, respectively. The expression is obtained under the assumption that the electronic and magnetic densities in the tip vary slowly with energy. While this conductance contribution relates the DOS and magnetic density in the tip ($n(\dote{F}-eV)$, $\bfm(\dote{F}-eV)$) and substrate ($N(\bfr,\dote{F})$, $\bfM(\bfr,\dote{F})$), it is independent of the local spin moments.

\subsection{Dynamical conductance}
We next consider the expression for the first order conductance in the total local spin moment. This is provided by a similar consideration of the second contribution to the tunneling current, i.e. beginning from
\begin{align}
I_1(\bfr,t,V)=&
	\frac{2eT_0}{\hbar}\re\sum_{\bfp\bfk\sigma}\sigma_{\sigma\sigma}^z\sum_n
		T_1(\bfr-\bfr_n)
		\int_{-\infty}^t
			[f(\leade{\bfp})
\nonumber\\&\times\vphantom{\int}
			f(-\leade{\bfk})
			\av{S_n^z(t')}
			-f(-\leade{\bfp})
			f(\leade{\bfk})
\nonumber\\&\times\vphantom{\int}
			\av{S_n^z(t)}]
			e^{i(\leade{\bfp}-\leade{\bfk}+eV)(t-t')}
		dt'.
\end{align}
Here, we again have assumed that scattering between different momentum and spin channels within the tip and substrate is negligible. We replace the time-dependent spin by its Fourier transform, i.e. $\av{S_n^z(t)}=\int\av{S_n^z(\omega)}e^{i\omega t}d\omega/(2\pi)$. Along with the replacement of the momentum summations by energy integration over the corresponding spin resolved DOS of the tip and substrate, we find that this current can be written
\begin{align}
I_1(\bfr,t,V)=&
	\frac{2\pi eT_0}{h}\sum_n
		T_1(\bfr-\bfr_n)
		\int
		\av{S_n^z(\omega)}
		\cos\omega t
\nonumber\\&\times\vphantom{\int}
		[n(\dote{})M_z(\bfr,\dote{}')\cos\theta+m_z(\dote{})N(\bfr,\dote{}')]
\nonumber\\&\times\vphantom{\int}
		[
			f(\dote{})
			f(-\dote{}')
			\delta(\dote{}-\dote{}'-\omega+eV)
\nonumber\\&\vphantom{\int}
			-f(-\dote{})
			f(\dote{}')
			\delta(\dote{}-\dote{}'+eV)
		]
		d\dote{}d\dote{}'\frac{d\omega}{2\pi}.
\end{align}
We, thus, obtain the conductance
\begin{align}
\frac{\partial I_1(\bfr,t,V)}{\partial V}=&
	\frac{\pi^2\sigma_0T_0}{4k_BT}\sum_nT_1(\bfr-\bfr_n)\int
		\av{S^z_n(\omega)}\cos(\omega t)
\nonumber\\&\times\vphantom{\int}
		[
		m_z(\dote{})
		N(\bfr,\dote{}')
		+n(\dote{})
		M_z(\bfr,\dote{}')
		\cos(\theta)]
\nonumber\\&\times\vphantom{\int}
		[\delta(\dote{}-\dote{}'-\omega+eV)
		+\delta(\dote{}-\dote{}'+eV)]
\nonumber\\&\times\vphantom{\int}
	\cosh^{-2}\frac{\dote{}}{2k_BT}
	d\dote{}d\dote{}'\frac{d\omega}{2\pi}.
\nonumber\\\rightarrow&
	\pi^2\sigma_0T_0T_1\sum_n\int
		\av{S^z_n(\omega)}\cos(\omega t)
		\Bigl[
		m_z(\dote{})
\nonumber\\&\times\vphantom{\int}
		N(\bfr,\dote{F})
		+n(\dote{})
		M_z(\bfr,\dote{F})
		\cos(\theta)\Bigr]
\nonumber\\&\times\vphantom{\int}
		[\delta(\dote{}+eV-\omega)
		+\delta(\dote{}+eV)]
	d\dote{}\frac{d\omega}{2\pi},
\label{eq-dI1}
\end{align}
as $	T\rightarrow0$.

\subsection{Conductance with spin-spin correlations}
The last term contained in the original expression given in Eq. (\ref{eq-I}) can be written as
\begin{align}
I_2(\bfr,t,V)=&
	\frac{2e}{\hbar}\re\sum_{\sigma\sigma'}\sum_{nm}T_1(\bfr-\bfr_n)T_1(\bfr-\bfr_m)
		\int
			n_\sigma(\dote{})
\nonumber\\&\vphantom{\int}\times
			N_{\sigma'}(\dote{}')
		\int_{-\infty}^t
			e^{i(\dote{}-\dote{}'+eV)(t-t')}
		[
			f(\dote{})
			f(-\dote{}')
\nonumber\\&\vphantom{\int}\times
			\bfsigma_{\sigma\sigma'}
			\cdot\av{\bfS_n(t)\bfS_m(t')}\cdot
			\bfsigma_{\sigma'\sigma}
			-f(-\dote{})f(\dote{}')
\nonumber\\&\vphantom{\int}\times
			\bfsigma_{\sigma'\sigma}
			\cdot\av{\bfS_m(t')\bfS_n(t)}\cdot
			\bfsigma_{\sigma\sigma'}
		]
		dt'd\dote{}d\dote{}',
\end{align}
explicitly expressed in terms of the spin-spin correlation functions of the local spins. Although the spin-spin correlation function e.g. $\av{\bfS_n(t)\bfS_m(t')}$ provides all information from the local spin correlations to the tunneling current, it is convenient to rewrite this general function in terms of the propagators $\chi^{\pm\mp}_{nm}(t,t')=(-i)\av{S^\pm_n(t)S_m^\mp(t')}$ and $\chi^z_{nm}(t,t')=(-i)\av{S^z_n(t)S_m^z(t')}$. We notice in the stationary regime, that these correlation functions depends on the time-difference $t-t'$, which allow us to express them through the Fourier transforms, e.g. $\chi_{nm}^z(t,t')=\int\chi_{nm}^z(\omega)e^{-i\omega(t-t')}d\omega/(2\pi)$. These remarks lead to that we can write the third contribution to the tunneling current according to
\begin{widetext}
\begin{align}
I_2(\bfr,t,V)=&
	\frac{i\pi e}{8\hbar}\sum_{nm}T_1(\bfr-\bfr_n)T_1(\bfr-\bfr_m)
		\int
			\Bigl(
				f(\dote{})f(-\dote{}')\delta(\dote{}-\dote{}'-\omega+eV)
				-f(-\dote{})f(\dote{}')\delta(\dote{}-\dote{}'+\omega+eV)
			\Bigr)
\nonumber\\&\vphantom{\int}\times
		\Bigl(
				[\chi^{+-}_{nm}(\omega)+\chi^{-+}_{nm}(\omega)]
		[
			n(\dote{})N(\dote{}')
			-\bfm(\dote{})\cdot\bfM(\dote{}')
		]
\nonumber\\&\vphantom{\int}
		+
			[\chi^{+-}_{nm}(\omega)-\chi^{-+}_{nm}(\omega)]
		[
			n(\dote{})M_z(\dote{}')\cos\theta-m_z(\dote{})N(\dote{}')
		]
\nonumber\\&\vphantom{\int}
		+4\chi^z_{nm}(\omega)
			[
				n(\dote{})N(\dote{}')
				+\bfm(\dote{})\cdot\bfM(\dote{}')
			]
		\Bigr)
		d\dote{}d\dote{}'\frac{d\omega}{2\pi}.
\end{align}
Despite the unappealing length of this expression, it provides a convenient starting point for analyses of spin inelastic tunneling spectroscopy. The conductance corresponding to this tunneling current becomes\cite{franssonNL2009}
\begin{align}
\frac{\partial I_2(\bfr,t,V)}{\partial V}=&
	i\biggl(\frac{\pi}{4}\biggr)^2\frac{\sigma_0}{2k_BT}\sum_{nm}T_1(\bfr-\bfr_n)T_1(\bfr-\bfr_m)
	\int
		\Bigl(
			f(\dote{})\delta(\dote{}-\dote{}'-\omega+eV)
			+
			f(-\dote{})\delta(\dote{}-\dote{}'+\omega+eV)
		\Bigr)
\nonumber\\&\vphantom{\int}\times
		\Bigl(
				[\chi^{+-}_{nm}(\omega)+\chi^{-+}_{nm}(\omega)]
		[
			n(\dote{})N(\dote{}')
			-\bfm(\dote{})\cdot\bfM(\dote{}')
		]
\nonumber\\&\vphantom{\int}
		+
			[\chi^{+-}_{nm}(\omega)-\chi^{-+}_{nm}(\omega)]
		[
			n(\dote{})M_z(\dote{}')\cos\theta-m_z(\dote{})N(\dote{}')
		]
\nonumber\\&\vphantom{\int}
		+4\chi^z_{nm}(\omega)
			[
				n(\dote{F})N(\dote{})
				+\bfm(\dote{})\cdot\bfM(\dote{}')
			]
		\Bigr)
		\cosh^{-2}\frac{\dote{}'}{2k_BT}
		d\dote{}d\dote{}'\frac{d\omega}{2\pi}
		\rightarrow
\nonumber\\\rightarrow&
	i\biggl(\frac{\pi}{2}\biggr)^2\frac{\sigma_0}{2}\sum_{nm}T_1(\bfr-\bfr_n)T_1(\bfr-\bfr_m)
	\int
		\Bigl(
			f(\dote{})\delta(\dote{}-\omega+eV)
			+
			f(-\dote{})\delta(\dote{}+\omega+eV)
		\Bigr)
\nonumber\\&\vphantom{\int}\times
		\Bigl(
				[\chi^{+-}_{nm}(\omega)+\chi^{-+}_{nm}(\omega)]
		[
			n(\dote{})N(\dote{F})
			-\bfm(\dote{})\cdot\bfM(\dote{F})
		]
\nonumber\\&\vphantom{\int}
		+
			[\chi^{+-}_{nm}(\omega)-\chi^{-+}_{nm}(\omega)]
		[
			n(\dote{})M_z(\dote{F})\cos\theta-m_z(\dote{})N(\dote{F})
		]
\nonumber\\&\vphantom{\int}
		+4\chi^z_{nm}(\omega)
			[
				n(\dote{})N(\dote{F})
				+\bfm(\dote{})\cdot\bfM(\dote{F})
			]
		\Bigr)
		d\dote{}\frac{d\omega}{2\pi},\
		T\rightarrow0.
\label{eq-dI2}
\end{align}
\end{widetext}

We summarize this section by pointing out that Eqs. (\ref{eq-dI0}), (\ref{eq-dI1}), and (\ref{eq-dI2}) serve as the main results of this paper. Below, we shall clarify the use of the partitioning of the conductance and how we can interpret experimental results in terms of the different contributions.

We notice, however, that the electronic structures in the tip and substrate are described in a simple fashion, in that we assume dependence on a single momentum vector in terms of the homogeneous GFs, e.g. $g_{\bfk\sigma}(t,t')$ for the substrate electrons. Beginning from the current given in Eq. (\ref{eq-I}) we can, however, repeat the above derivation allowing for an inhomogeneous description of the electronic structures, using e.g. $g_{\sigma\sigma'}(\bfk,\bfk';t,t')$ for the substrate electrons. This will modify the formulas given for the conductance, something which goes beyond the scope of the present study.

In connection to this, we also notice that we, here, have disregarded that the electronic and magnetic densities in the substrate are influenced by the local spin moments of the adatoms. In reality, the local spins and the substrate mutually influence one another and self-consistently create a total effective magnetic moment locally around the adatoms. The basic formulas, Eqs. (\ref{eq-dI0}), (\ref{eq-dI1}), and (\ref{eq-dI2}), remain unchanged, however, by a self-consistent treatment of the substrate and adatoms. From the point of view we take in this paper, we can assume that the self-consistently created magnetic moment in the substrate is already included in the density $\bfM$.

\section{Discussion of the tunneling conductance}
\label{sec-analysis}
The first contribution to the conductance, see Eq. (\ref{eq-dI0}), essentially captures the Tersoff-Hamann theory\cite{tersoff1983,reittu1997,wortmann2001,heinze2006} for magnetic tip and substrate. One should notice, however, that we here have neglected the specific relation between the local tunneling matrix element $T_0$ and the local DOS in the substrate, since we here are interested in the possibility to resolve its dependence on local spin moments located on the substrate. In studies of magnetic surfaces, this conductance provides a sufficient tool for analysis of the magnetic structure of the surface, by using the spin-polarization of the electronic structure at the Fermi level, since the density $\bfM(\dote{F})$ contains all such information. An analysis of local spin moments adsorbed onto the substrate can, in this model, only be performed on the level of the DOS and spin splitting of electron states at an energy defined by the Fermi level and the applied bias voltage. The total spin moment of the adatom can, nevertheless, not be accessed through this conductance. This fact is based on that although one can include the adsorbate into the electronic and magnetic densities of the surface, one only obtains energy dependent information of the adsorbate, while the spin moment is defined by the spin-polarization integrated over all energies. Any deeper information about the adsorbate can, thus, not be achieved. In this sense, we can consider the contribution $\partial I_0/\partial V$ as generating a back-ground conductance which includes features originating from the substrate.

The second conductance contribution, see Eq. (\ref{eq-dI1}), contains terms which depend on the spin-polarization in the tip, $m_z(\dote{})$, and of the substrate, $M_z(\bfr,\dote{F})$, respectively, and vanishes in case both the tip and substrate are non-magnetic. An important result here is that this contribution explicitly identifies a linear relationship between local spin moment, $\av{S_n^z(\omega)}=\int\av{S_n^z(t)}e^{-i\omega t}dt$, and the tunneling conductance. Eq. (\ref{eq-dI1}) implies that one, in principle, can obtain a direct estimation of the size of the local spin moment simply by measuring the differential conductance at its location. In order to achieve this functionality, it is necessary, as seen in Eq. (\ref{eq-dI1}), that the spin-polarizations of the tip and substrate are known. We further discuss this application below. It is worth pointing out, however, that although we have assumed stationary, i.e. time-independent, conditions, there is nevertheless a time-dependent component in the tunneling conductance. This time-dependence is generated by the dynamics of the local spin and it is this feature that would enable a read-out of the local spin moment. The occurrence of the time-dependent component in the conductance, and tunneling current, also suggests a mechanism which may be employed for generation of high frequency electrical currents. The effective magnetic fields acting on the local spins, e.g. anisotropy fields etc., may be of the order of meV, see below, which would enable generation of high GHz to THz ac currents.

The last conductance contribution, see Eq. (\ref{eq-dI2}), provides signatures that are generated by the spin-spin correlations, or spin fluctuations, occurring in the adsorbate. It is important to note that this conductance is finite for any polarization of the tip and the substrate, even when both electrodes are non-magnetic. Hence, regardless of the electronic and magnetic conditions of the system, this conductance directly depends on spin fluctuations in the adsorbate. It is also noticeable that the formulation of $\partial I_2/\partial V$ suggests one, in principle, will be able to study and distinguish between particular spin excitations in the adsorbate. This functionality is expressed from the fact that the sum and difference of the correlation functions $\chi_{nm}^{+-}$ and $\chi_{nm}^{-+}$ are multiplied by different combinations of the electronic and magnetic structures of the tip and substrate. One is therefore capable to configure the STM set-up in order to probe particular spin excitations in the local spin. This will be discussed further below.

\section{Physical information contained in the tunneling conductance}
\label{sec-exp}
We now discuss a few different physical examples which are introduced in order to shine some light on different aspects of the tunneling conductance. Before we go into the examples, however, we introduce the model of the local spins adsorbed on the surface.

We consider a cluster of spins on the substrate and write the Hamiltonian for the spin $\bfS_n$ in the cluster according to $\Hamil_n=g\mu_B\bfB\cdot\bfS_n$, where $\bfB$ is an external magnetic field. The effective exchange interaction between the spin moments in the cluster is given by a Heisenberg model $\Hamil_J=-J\sum_{n\neq m}\bfS_n\cdot\bfS_m$. This effective exchange comprise a combination of e.g. direct Heisenberg exchange and RKKY-like (Ruderman-Kittel-Kasuya-Yosida) exchange. The sign of the effective $J$ may, thus, vary with distance between the spins in the cluster.\cite{meier2008} In this way we describe clusters of spins with a total spin moment $\bfS$. We define the resulting eigensystem $\{\dote{\sigma},\ket{S,\sigma}\}$, $\sigma=-S,-S+1,\ldots,S$, for the eigenenergies and eigenstates, respectively, of the model 
\begin{align}
\Hamil_S=&
	\sum_n\Hamil_n+\Hamil_J+\sum_n[D(S_n^z)^2+E\{(S_n^x)^2-(S_n^y)^2\}],
\label{eq-spinmodel}
\end{align}
where we have added the spin anisotropy fields $D$ and $E$. This model is pertinent to the recent studies of transition metal elements adsorbed onto surfaces e.g. Fe and Mn on CuN surfacem,\cite{hirjibehedin2006,hirjibehedin2007} and Fe and Co on Pt surface.\cite{balashov2009} In those elements, the magnetism is constituted by the electrons in the $d$-shell whereas the electrons in the $s$- and $p$-shell strongly hybridize with the surface states in the substrate. We, therefore, assume that the $s$- and $p$-electrons in the adsorbate are included in the Hamiltonian for the surface electrons and, below, we shall refer to them as conduction electrons. The electrons in the $d$-shell can be regarded as only weakly hybridizing with the conduction electrons which, therefore, enables us to disregard their contribution to the conductance.

Under the assumption of very weak coupling between the $d$-electrons and the conduction electrons, which is pertinent to the recent experiments on e.g. Fe and Mn on CuN,\cite{hirjibehedin2006,hirjibehedin2007} we write the total dynamical local spin moment in terms of the eigensystem defined for the cluster according to
\begin{align}
\av{S^z(\omega)}=2\pi\sum_{\sigma=-S}^S\sigma f(\dote{\sigma})\delta(\omega-\dote{\sigma}),
\end{align}
where $f(\dote{\sigma})$ provides the occupation of the state $\ket{S,\sigma}$. Hence, the conductance is proportional to $\int\av{S^z(\omega)}d\omega/(2\pi)=\sum_\sigma\sigma f(\dote{\sigma})$, c.f. Eq. (\ref{eq-dI1p}) below.

In order to also describe the experiments conducted on e.g. a Pt surface,\cite{balashov2009} we phenomenologically introduce a width of the local spin excitations by replacing Dirac delta functions by Lorentzian functions. Such a treatment can be physically motivated by that this would be the essential effect of a mean field approximated dressing of the bare spin averages and spin-spin correlation functions. This phenomenological model is sufficient for our present purposes since we do not attempt to explain all the details of the experiments.

\subsection{Estimating the spin moment}
\label{ssec-est}
The conductance in Eq. (\ref{eq-dI1}) can be directly linked to the total local spin moment $\av{S_n^z}$. This is simplest seen by assuming that the electronic and magnetic densities in the tip and substrate vary slowly with the energy. Calculating the Fourier transform of $\partial I_1(t)/\partial V$ and integrating over all frequencies gives
\begin{align}
\frac{\partial I_1(\bfr,V)}{\partial V}=&
	2\pi^2\sigma_0T_0
	\sum_nT_1(\bfr-\bfr_n)
	\av{S_n^z}
		\Bigl[
		m_z(\dote{F}-eV)
\nonumber\\&\times
		N(\bfr,\dote{F})
		+n(\dote{F}-eV)
		M_z(\bfr,\dote{F})
		\cos(\theta)\Bigr],
\label{eq-dI1p}
\end{align}
where we have identified the total spin moment of the $n$th spin by $\av{S_n^z}=\int\av{S^z_n(\omega)}d\omega/(2\pi)$. In this fashion, we obtain a linear relationship between the differential conductance $\partial I/\partial V$ and the average spin moment $\av{S_n^z}$.

Our theory suggests that the STM conductance generates a time-dependent signal, in agreement with earlier studies which suggest time-dependent noise spectroscopy using STM tunneling current,\cite{balatsky2002,nussinov2003} regardless of the  time-dependence, or time-independence, of the bias voltage. The Fourier transform of this signal provides an energy resolved signal from which additional understanding, i.e. dynamics, about the spin systems can be extracted.

Consider the simple example, given for e.g. a single ($n=1$) spin moment of $S=1$, for which $\sigma=0,\pm1$. The anisotropy fields $D$ and $E$ break up the spin symmetry, such that the spin levels are given by $\dote{0}=0$ and $\dote{\pm1}=D\pm\sqrt{E^2+(g\mu_BB)^2}$, for $\bfB=B\hat{\bf z}$. The dynamical spin moment, thus, becomes $\av{S^z(\omega)}=\pi[f(\dote{1})\delta(\omega-\dote{1})-f(\dote{-1})\delta(\omega-\dote{-1})]$, which generates a non-vanishing current when the populations $f(\dote{\pm1})$ of the states $\ket{1,\pm1}$ are different. This observation holds for an adatom with any value of the spin and emphasizes the fact that the local spin has to have a definite moment in order to be measurable through the SP-STM. Using this dynamical spin moment in Eq. (\ref{eq-dI1}) and assuming slowly varying electronic and magnetic densities in the tip and substrate, results for low temperatures in
\begin{align}
\label{eq-dI1est}
\frac{\partial I_1(t)}{\partial V}\sim&
	f(\dote{1})\cos\dote{1}t
	-f(\dote{-1})\cos\dote{-1}t.
\end{align}

More generally, the time-dependence of the conductance can be written
\begin{align}
\frac{\partial I_1(t)}{\partial V}\sim&
	\sum_\sigma\sigma f(\dote{\sigma})\cos\dote{\sigma}t.
\end{align}

\begin{figure}[t]
\begin{center}
\includegraphics[width=0.45\textwidth]{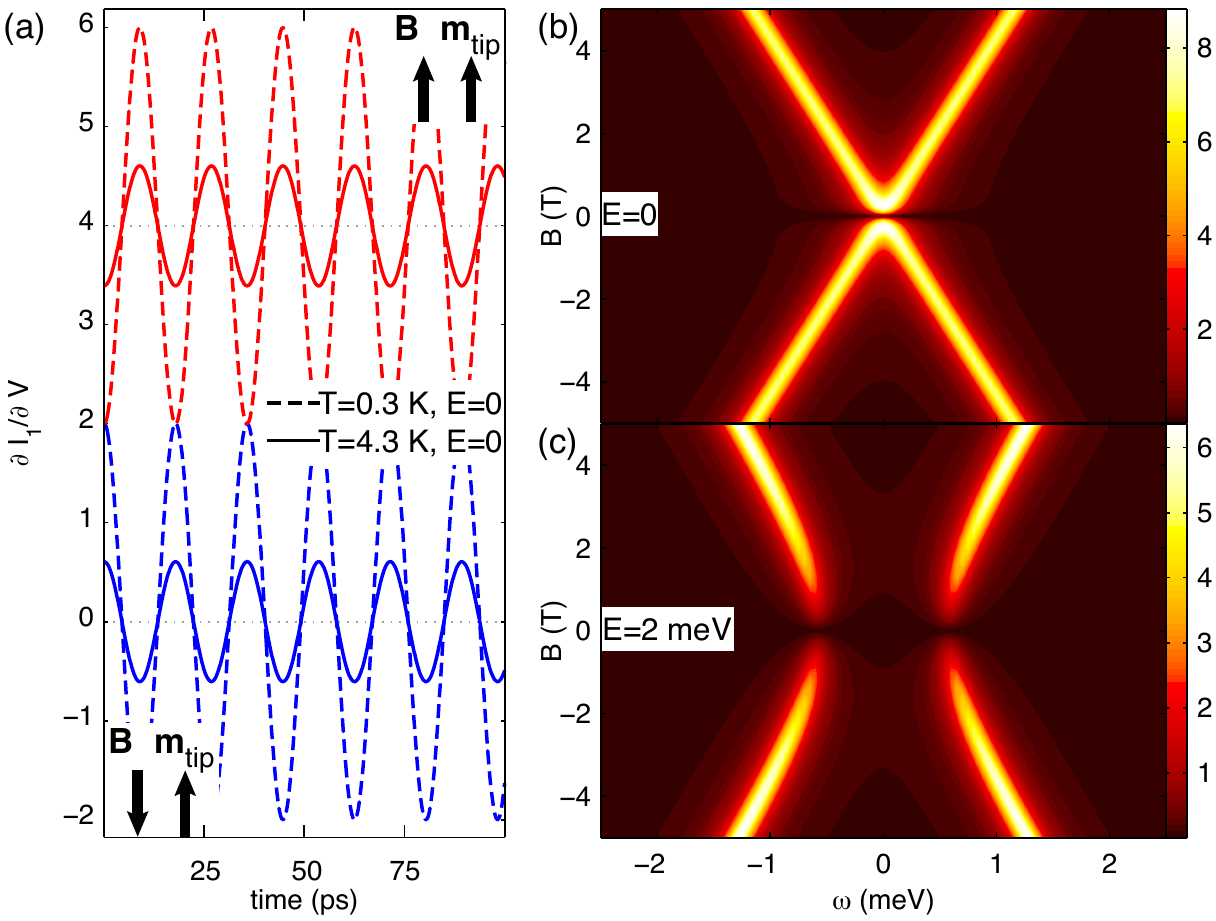}
\end{center}
\caption{(Color online) Various aspects of $\partial I_1/\partial V$ for a single $S=1$ adatom. (a) $\partial I_1(t,V)/\partial V$, Eq. (\ref{eq-dI1}), for $\bfB=B\hat{\bf z}$ and $\bfm_{\tip}$ anti-parallel, $B=-1$ T (lower plots), and parallel, $B=1$ T (upper plots). The plots correspond to $T=0.3$ K, $E=0$ (dashed), $T=4.3$ K, $E=0$ (solid). (b) and (c) $|\partial I_1(\omega,V,B)/\partial V|$ as function of $\omega$ (horizontal axis) and $B$ (vertical axis) at $T=0.3$ K, and (b) $E=0$ and  (c) $E=2$ meV. Bright (dark) colors correspond to large (small) amplitude. The conductances have been normalized by $2\pi^2\sigma_0T_0T_1[m_zN+nM_z]$. Here, $D=-10$ meV, $g=2$, $m_z=3n(\dote{F})/4$, $M_z=0$, and $V=0.3$ meV}
\label{fig-IoftwB}
\end{figure}

The period of the conductance oscillations are, thus, directly linked to the energy levels of the spin states, see Fig. \ref{fig-IoftwB} (a) for an example of the time-dependent conductance for two different set-ups of the SP-STM. The period of the conductance oscillations can, thus, be changed by applying an external magnetic field $B_z$ in order to vary the energy levels $\dote{\sigma}$, as is illustrated in Fig. \ref{fig-IoftwB} (b), showing the magnetic field dependence of the Fourier transformed dynamical conductance. The time scale associated with the conductance oscillations is given by the set of eigenenergies $\dote{\sigma}$, which means that only very low lying excitation energies (0.066 meV corresponding to 100 GHz) are reachable by means of the state-of-the-art experimental technology. Although detection of spin moments using this method definitely challenges todays experimental resources and capabilities, it should be an accessible regimes within the nearest future. 

It is important to notice, however, that the presence of the time-dependent component in the conductance opens the possibility to generate high frequency ac currents. Due to the high anisotropy fields acting on the local spin moment, of the order to meV (see Sec. \ref{ssec-anis}), one would be able to generate electrical and spin-dependent ac currents and/or ac voltages with frequencies in the THz regime.

\subsection{Anisotropy parameters}
\label{ssec-anis}
Our next example of the usefulness of Eq. (\ref{eq-dI1p}) is to consider an $S=1$ adatom, and we plot for this system the conductance $\partial I_1/\partial V$, see Fig. \ref{fig-IoftwB} (a), where the anisotropy fields $D=-10$ meV and $E=0$ meV for different temperatures, i.e. T=0.3 K (dashed) and T=4.3 K (solid). These parameters are pertinent to the recent studies of Co/Pt(111).\cite{gambardella2003,balashov2009,meier2008} The plots in (a) clearly show the phase shift of $\pi$ when the magnetic field $\bfB$ is reversed from anti-parallel (lower) to parallel (upper) orientation with $\bfm_{\tip}$. The magnetization curves are S-shaped (not shown), which correspond to a paramagnetic behavior of the local spin and which was observed in experiments.\cite{meier2008}

From the dynamical conductance, $\partial I(t)/\partial V$ or $\partial I(\omega)/\partial V$, one can obtain information about the magnetic anisotropy fields $D$ and $E$ acting on the local spin. In the case of $D<0$ and $E=0$, the spin moment points perpendicular to the surface, which generates a maximal conductance. The response remains strong for values of $B$ sufficiently large to maintain a spin-splitting of the adatom which is larger than the thermal excitation energy $k_BT$. This is illustrated in Fig. \ref{fig-IoftwB} (b), where $|\partial_1I(\omega,B)/\partial V|$ is plotted for an $S=1$ adatom, as function of $\omega$ and $B$. Bright (dark) colors correspond to large (small) amplitude of $|\partial_1I(\omega,B)/\partial V|$. The lines that represent the ground state of the adatom cross at zero magnetic field, which is expected since the ground state is spin-degenerate at $B=0$.

A finite value of $E$, on the other hand, provides a tilting of the spin moment away from the perpendicular orientation, which leads to a weakened conductance. This is illustrated in Fig. \ref{fig-IoftwB} (c), where $E=2$. The finiteness of $E$ breaks the spin-degeneracy of the ground state at vanishing magnetic field. The ground state is, however, pointing parallel to the substrate, which leads to that $\av{S^z(\omega)}=0$ at $B=0$. The resulting magnetization line remains S-shaped (not shown). Despite the low temperature, however, the magnetization curve is reminiscent of the magnetization curve for higher temperature. This is expected since the anisotropy field $E$ introduces an energy barrier for the adatom to overcome in order to point its spin perpendicular to the substrate.

In the production of the plots in Fig. \ref{fig-IoftwB} (b) and (c) we replaced the Dirac delta function by a Lorentzian function with a broadening of 0.01 meV. This broadening is roughly 2 order of magnitudes smaller than what is expected from the experimental set-up with Co/Pt(111), see Sec. \ref{ssec-fluct} for estimates, and is chosen in order to emphasize the different behavior of the conductance of different anisotropy parameters. Using more realistic anisotropy parameters blurs the resulting image, however, the main difference between the character of the conductance for vanishing and finite field $E$ can be resolved.

\subsection{Character of exchange interaction parameter}
\label{ssec-exJ}
Studies of the dynamical conductance can also be used to reveal the sign of the effective exchange interaction between magnetic adatoms located on the surface. Consider for example a spin dimer. In case of ferromagnetic exchange, $J>0$, the ground state of the spin dimer is a spin triplet. Varying the magnetic field, the ground state of the dimer acquires a magnetic moment e.g. $\av{S^z}=S$ or $-S$, depending on whether the magnetic field is anti-parallel or parallel with the magnetic moment in the tip, in analogy with the single spin case discussed above. Hence, there is a measurable dynamical conductance. In case of an anti-ferromagnetic exchange, however, the ground state of the spin dimer is a spin singlet, with zero magnetic moment. Then, the dynamical conductance vanishes, c.f. Eq. (\ref{eq-dI1}), and the total conductance is strictly time-independent, except for possible noise fluctuations.

\subsection{Spin fluctuations}
\label{ssec-fluct}

To illustrate the effect of spin fluctuactions, we consider a local spin moment $\bfS$ comprising two coupled spins $\bfS_n$, $n=1,2$, a spin dimer, and consider them to be anti-ferromagnetically coupled. The ground state is a spin singlet $\ket{S=0,\sigma=0}$, while the first excited states constitute a spin triplet $\ket{S=S_1+S_2,\sigma=0,\pm(S_1+S_2)}$. Assuming an exchange energy $|J|=|E_S-E_T|>k_BT$, where $E_{T(S)}$ denotes the triplet (singlet) energy, in order to prevent thermal excitations at zero bias, the equilibrium conductance is given by the elastic tunneling between the tip and the substrate only, i.e. $dI/dV=dI_0/dV$. Effects from tunneling electrons scattering off the local spin moment averages to zero. 

\begin{figure*}[t]
\begin{center}
\includegraphics[width=0.75\textwidth]{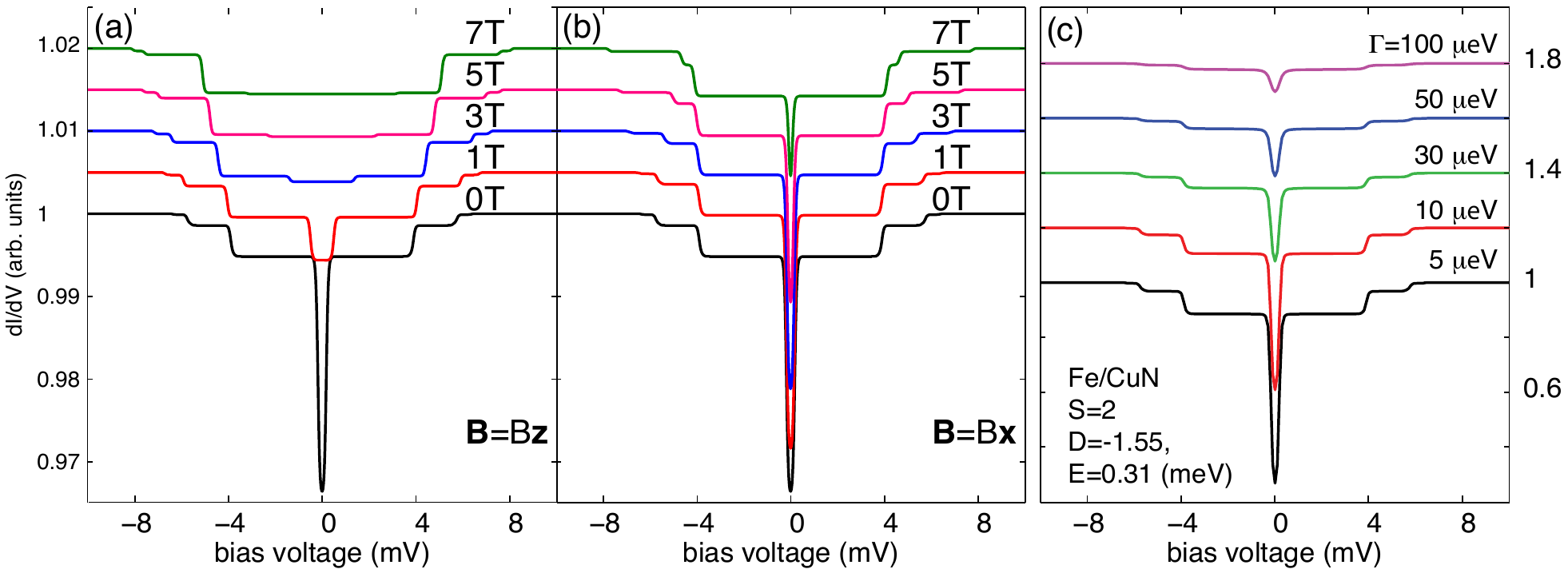}
\end{center}
\caption{(Color online) $dI/dV$ for Fe/CuN under different external magnetic fields at $T=0.5$ K; (a) $\bfB=B{\bf z}$, (b) $\bfB=B{\bf x}$, and (c) for different $\Gamma$. Plots are off-set for clarity.}
\label{fig-S2BzBx}
\end{figure*}

The coupling to the tunneling electrons via the spin-spin interaction e.g. $\cdagger{\bfp}\bfsigma_{\sigma\sigma'}\cdot\bfS_n\cs{\bfk\sigma}$ enables, on the other hand, each individual spin constituting $\bfS$ to undergo spin-flip transitions which are assisted by spin-flips of the tunneling electrons. Due to this coupling, the correlation function e.g. $\bfsigma_{\sigma\sigma'}\cdot\av{\bfS_n(t)\bfS_m(t')}\cdot\bfsigma_{\sigma''\sigma}$ is non-vanishing, in general. The spin-spin interaction, thus, provides a coupling between the singlet and triplet states which supports transitions between them. As a result of these transitions, a new channel for conductance opens at bias voltages $V\geq|J|/e$.

The spin-spin correlation functions $\chi^{-+}_{nm}(\omega)$, $\chi^{+-}_{mn}(\omega)$, and $\chi^z_{nm}(\omega)$ are calculated in terms of the eigensystem for the total spin, giving
\begin{subequations}
\label{eq-chi}
\begin{align}
	\chi_{nm}^{\mp\pm}(\omega)=&
	(-i)2\pi\sum_{iv}
		\bra{i}S_n^\mp\ket{v}\bra{v}S_m^\pm\ket{i}
		P(E_i)
\nonumber\\&\times
		[1-P(E_v)]\delta(\omega+E_i-E_v),
\label{eq-chimp}\\
	\chi_{nm}^z(\omega)=&
	(-i)2\pi\sum_{iv}
		\bra{i}S_n^z\ket{v}\bra{v}S_m^z\ket{i}
		P(E_i)
\nonumber\\&\times
		[1-P(E_v)]\delta(\omega+E_i-E_v),
\label{eq-chiz}
\end{align}
\end{subequations}
where $i\ (v)$ denotes the initial (intermediate) state, whereas $P(x)$ accounts for the population in the corresponding state. Eq. (\ref{eq-chi}) provides the general qualitative features of the spin-spin correlation function, and shows that there will appear steps in $dI/dV$ whenever the bias voltage matches the transition energy $\pm(E_i-E_v)$. In the case of a spin dimer, for instance, there appear steps in $dI/dV$ when the bias voltage supports the inelastic transitions between the single and triplet states. In case of e.g. Fe/CuN, the calculated results are plotted in Fig. \ref{fig-S2BzBx}, for a non-magnetic tip and substrate, showing that the presented theory reproduces the results discussed in Ref. \onlinecite{rossier2009} and shows an excellent agreement with experiments.\cite{hirjibehedin2006,hirjibehedin2007}

\begin{table}[b]
\caption{Anisotropy parameters, $D$ and $E$, used for the conductance plots given in  Fig. \ref{fig-S2BzBx} (c) and \ref{fig-FeCo}, the best widths, $\Gamma$, for the intermediate states in the spin-spin correlation functions.}
\label{tab-anis}
\begin{tabular}{ccccc}
\hline\hline
& S & D (meV) & E (meV) & $\Gamma$ (meV) \\\hline
Fe/CuN & 2 & -1.55 & 0.31 & $0.01 - 0.03$ \\
Fe/Pt(111) & 3/2 & -3.25 & 0 & $5$ \\
Co/Pt(111) & 1 & -10 & 2 & $10$ \\
\hline\hline
\end{tabular}
\end{table}

\begin{figure}[b]
\begin{center}
\includegraphics[width=8.5 cm]{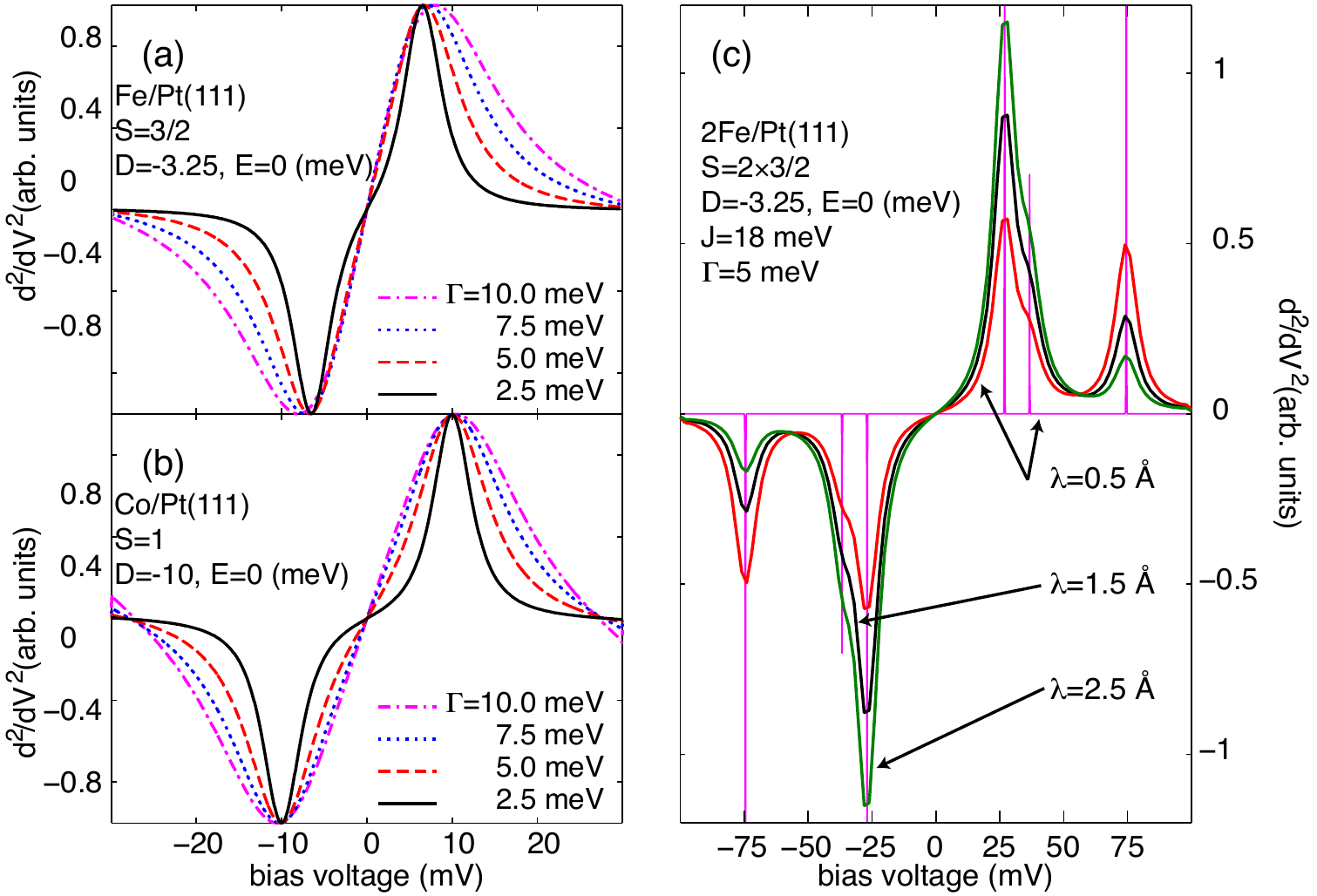}
\end{center}
\caption{(Color online) $d^2I/dV^2$ for (a) Fe/Pt(111) and (b) Co/Pt(111) for different widths $\Gamma$, and (c) 2Fe/Pt(111) for different decay lengths $\lambda$, at $T=4.3$ K. In panel (c), $d^2I/dV^2$ with zero broadening and $\lambda=0.5$ \AA\ is plotted for reference. }
\label{fig-FeCo}
\end{figure}

Signatures of the excitations in the experimental measurements do have a finite width, which corresponds to that the intermediate states have finite lifetimes. Replacing the delta functions in Eq. (\ref{eq-chi}) by Lorentzian funtions $1/(x^2+\Gamma^2)$, we phenomenologically include the (uniform) lifetime $\hbar/\Gamma$ for all intermediate states. In case of a single Fe on CuN with spin $S=2$,\cite{hirjibehedin2007} and the anisotropy parameters given in Tab. \ref{tab-anis}, we find that the Fe spin is weakly coupled ($\Gamma\sim10 - 30\ \mu$eV) to the Cu(100) through the CuN layer, see  \ref{fig-S2BzBx} (c). This value is extracted by comparing the ratio between the maximal and minimal conductance with the experimental result ($\sim 1/2$). Similarly, we also extract the widths of the single Fe $(S=3/2)$ and Co $(S=1)$ adsorbed onto Pt(111) surface.\cite{balashov2009} In Fig. \ref{fig-FeCo} (a) and (b), we have plotted $d^2I/dV^2$ for Fe (upper) and Co (lower) and we find the best correspondence with experiments using the parameters given in \ref{tab-anis}. In Fig. \ref{fig-FeCo} (c), we finally provide a computation of the Fe dimer on Pt(111) reported in Ref. \onlinecite{balashov2009}. Because of the presence of two Fe atoms, we make use of space dependence of the tunneling rate and the decay length $\lambda$, and plot the computed $d^2I/dV^2$ for different values of the decay length. In comparison with the experimental results, the decay length of 0.5 \AA\ gives the best agreement, suggesting a very rapid spatial decay of the spin-dependent tunneling rate. In the figure, we for reference have also included the resulting $d^2I/dV^2$ in absence of the broadening.

In the calculations we have estimated the population numbers $P(E_{i(v)})$ of the states $\ket{i(v)}$ involved in the spin-spin correlation functions, c.f Eq. (\ref{eq-chi}), by making use of the following observation. By expanding the correlation function e.g. $\chi^{+-}_{nm}(t,t')$ in terms of its eigenstates, we can write it as $\chi_{nm}^{+-}(t,t')=\sum_{iv}\av{i|S_n^-|v}\av{v|S^+_m|i}(-i)\av{(\ddagger{i}\dc{v})(t)(\ddagger{v}\dc{i})(t')}$ where $\ddagger{i}$ ($\dc{i}$) creates (annihilates) a particle in the state $\ket{i}$. In the atomic limit, we can employ the decoupling $\av{(\ddagger{i}\dc{v})(t)(\ddagger{v}\dc{i})(t')}=\av{\ddagger{i}(t)\dc{i}(t')}\av{\dc{v}(t)\ddagger{v}(t')}=P(E_i)[1-P(E_v)]e^{i(E_i-E_v)(t-t')}$, where the population number $P(E_i)=\av{\ddagger{i}\dc{i}}$ can be estimated by using e.g. the Gibbs distribution $P(E)=e^{-\beta E}/\sum_ie^{-\beta E_i}$.

\subsection{Probing specific spin excitations}
\label{ssec-exc}
We observe in Eqs. (\ref{eq-dI2}) and (\ref{eq-chi}) that the transition sequences $\bra{i}S_n^+\ket{v}\bra{v}S_m^-\ket{i}$, $\bra{i}S_n^-\ket{v}\bra{v}S_m^+\ket{i}$, and $\bra{i}S_n^z\ket{v}\bra{v}S_m^z\ket{i}$ are associated with different projections of the spin-resolved LDOS in the tip and substrate. The sum and difference of the first two sequences couple to $n(\dote{})N(\dote{F})-\bfm(\dote{})\cdot\bfM(\dote{F})$ and $m_z(\dote{})N(\dote{F})-n(\dote{})M_z(\dote{F})\cos\theta$, respectively, while the last sequence couples to $n(\dote{})N(\dote{F})+\bfm(\dote{})\cdot\bfM(\dote{F})$. The different couplings reflect an ability to enhance or attenuate the response of certain inelastic transitions, at will, by using different combinations of electronic and magnetic densities in the tip and substrate.

\begin{figure}[t]
\begin{center}
\includegraphics[width=8.5 cm]{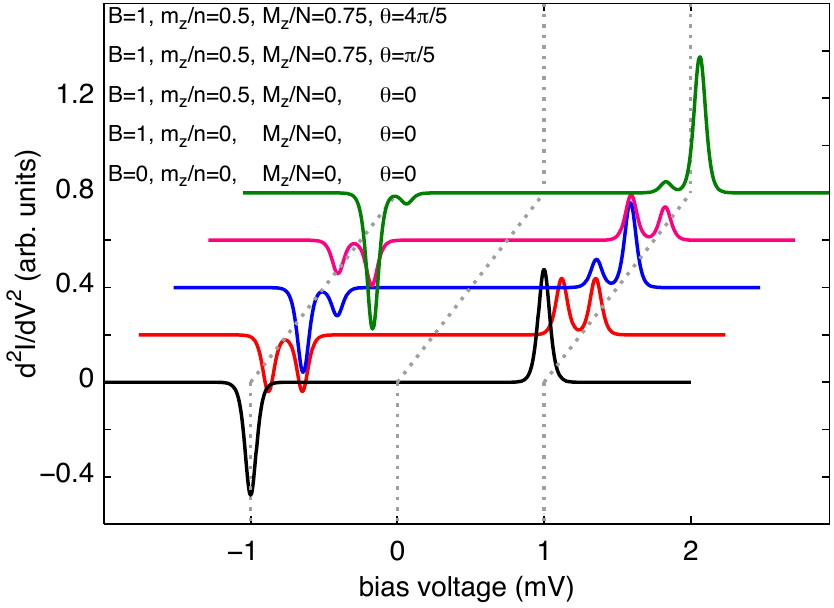}
\end{center}
\caption{(Color online) $d^2I/dV^2$ for a spin $S=1$ impurity, for various conditions of the external magnetic field $\bfB$, magnetizations of the tip ($\bfm$) and substrate ($\bfM$), and the relative angle ($\theta$) between the magnetization directions. Plots off-set for clarity.}
\label{fig-S1spstm}
\end{figure}

In STM without spin-polarization ($\bfm,\bfM=0$), for instance, the excitation spectrum can be analyzed to certain detail by means of applying an external magnetic field e.g. along the $z$-direction, of the spin quantization axis of the sample. Such application of the magnetic field introduces a Zeeman splitting of the levels, which thus leads to a separation of the peaks in the $d^2I/dV^2$. To be specific, consider a single adatom with $S=1$, and anisotropy fields $D<0$ and $E=0$, which is described by the states $\ket{S=1,\sigma=0,\pm1}$, where $E_\sigma<E_0$ at $\bfB=0$. In this system, the leading contribution to the term in $dI_2/dV$ containing $\chi_{nm}^{-+}(\omega)+\chi_{nm}^{+-}(\omega)$, c.f Eq. (\ref{eq-dI2}), is proportional to $\sum_{\sigma=\pm1}[P(E_\sigma)-P(E_0)][f(eV-E_\sigma+E_0)-f(eV+E_\sigma-E_0)]$, while the term containing $\chi_{nm}^{-+}(\omega)-\chi_{nm}^{+-}(\omega)$ vanishes. The former contribution generates steps in the conductance at bias voltages $eV=\pm(E_\sigma-E_0)$. Those steps are separated by $|E_{+1}-E_{-1}|$ whenever the states $\ket{1,0}$ and $\ket{1,\sigma}$ are unequally occupied. Hence, the spin splitting imposed by the external magnetic field is detectable through separate but equally high steps in the conductance. This splitting can be seen in Fig. \ref{fig-S1spstm}, where we plot $d^2I/dV^2$ of this scenario, where the lowermost curve correspond to non-magnetic conditions, whereas the second curve from below is obtained with $\bfB=B{\bf z}$ and $B=1$.

Using SP-STM opens further possibilities in the studies of spin systems, since then the term in $dI_2/dV$ containing $\chi_{nm}^{-+}(\omega)-\chi_{nm}^{+-}(\omega)$ is finite. In case of the $S=1$ adatom, the leading contribution to this term is proportional to $\sum_{\sigma=\pm1}\sigma[P(E_\sigma)+P(E_0)-2P(E_\sigma)P(E_0)][f(eV-E_\sigma+E_0)-f(eV+E_\sigma-E_0)]$, where it is important to notice that the contributions to this term have opposite signs. Hence, by combining the electronic and magnetic densities in the tip and substrate such that $m_zN-n\bfM_z\cos\theta<0$, this term leads to an attenuated (intensified) signal from the transition $\ket{1,0}\bra{1,+1}$ ($\ket{1,0}\bra{1,-1}$), c.f. third and fourth curves from below in Fig. \ref{fig-S1spstm}. By the same token, the signal is intensified (attenuated) when $m_zN-n\bfM_z\cos\theta>0$, see uppermost curve in Fig. \ref{fig-S1spstm}. While the details of the terms containing the sum and difference of the spin-spin correlation functions $\chi_{nm}^{-+}$ and $\chi_{nm}^{+-}$ differ from system to system, the general conclusion we can draw out of this observation is that the intensity of the signal from any specific transition depends on the magnetic densities of the tip and substrate, and on their relative orientation.

\subsection{Maximizing the signal from fluctuations}
\label{ssec-max}
Finally, we notice that the conductance $dI_0/dV$ vanishes for $nN+\bfm\cdot\bfM=0$, which corresponds to the case with a half-metallic tip and substrate such that their magnetic moments are in anti-parallel alignment. In this set-up also the conductance $dI_1/dV=0$, which implies that the measured signal is generated solely by $dI_2/dV$. As can be seen in Eq. (\ref{eq-dI2}), this conductance only depends on the transverse components, i.e. the sum and difference of $\chi_{nm}^{-+}$ and $\chi_{nm}^{+-}$, since the term containing $\chi_{nm}^z$ is proportional to $nN+\bfm\cdot\bfM\, (=0)$. Therefore, despite the presence of possible thermal noise, such a set-up would benefit from a very low current noise since most of the noise would be related to the spin fluctuations, that is, the noise we want to measure. This can be seen by identifying the spin-dependent current operator with $\delta \hat{I}(t)=T_1\bfS(t)\cdot\bfs$, where $\bfs=\sum_{\bfp\bfk\sigma\sigma'}\cdagger{\bfp}\bfsigma\cs{\bfk\sigma'}$. The current-current correlation function is then given by \cite{balatsky2002,nussinov2003} $\av{\delta\hat{I}(t)\delta\hat{I}(t')}=T_1^2\bfs\cdot\av{\bfS(t)\bfS(t')}\cdot\bfs$, where we average over the dynamics of the localized spins and over the ensemble of the tunneling electrons. Under the condition that $nN+\bfm\cdot\bfM=0$, the total dc current $I$ is  proportional to $T_1^2\int_{-\infty}^t\bfs\cdot\av{\bfS(t)\bfS(t')}\cdot\bfs\, d(t-t')$, and since the shot noise is approximately $\av{I_\text{shot}^2(\omega)}\sim I$,  the signal-to-noise ratio is about unity.

\section{Conclusions}
\label{sec-conclusion}
The theory we propose here for STM/STS measurements (essentially) contains the well-known Tersoff-Hamann term,\cite{tersoff1983,wortmann2001,heinze2006} Eq. (\ref{eq-dI0}), which maps the energy dependence electronic and magnetic densities of states, for energies defined by the Fermi level, of the substrate, and the applied bias voltage. For the situation where magnetic impurities are located on the substrate, our theory also contains a contribution which is proportional to the magnetic moment of the local spins, and allows for a quantitative analysis of spin moments using SP-STM. This is, hence, different from the currently most common analysis, which is solely focused on the energy dependent spin-polarization of the DOS. We show here that studies and analyses of both the dynamical conductance, $\partial I(V,t)/\partial V$ or $\partial I(V,\omega)/\partial V$, and the total conductance $\partial I(V)/\partial V$ can be linked to the dynamical, $\av{S^z(t)}$ or $\av{S^z(\omega)}$, and total, $\int\av{S^z(\omega)}d\omega/2\pi$, magnetic moments, respectively, of the local spins. This contribution generates different response at the local spins depending on their relative orientation compared to the spin moment on the SP-STM tip. To our knowledge, this contribution has not been discussed before, and the direct relation between the differential conductance and the local spin-moment is expected to make a significant impact on the potential capabilities in using SP-STM, and to extract quantitative information about local spin moments.

We also point out that despite the present difficulties to experimentally record the time-dependent component in the conductance in a high GHz or even THz regime, it is important to observe that the dynamical component to the conductance opens new possibilities for the generation of high frequency ac currents and/or voltages. Thus, by using local spin moments adsorbed onto metallic surface, such a scenario requires at least one magnetic electrode which produces a local magnetic field that can interact with the local spin moment and, thus, provide a time-dependent net contribution to the tunneling current, and conductance. Having in mind anisotropy fields from the substrate acting on the spin moment, it would be possible to generate spin-dependent ac currents/voltages in the THz regime, since the anisotropy fields may be of the order of meV.

It is important to note that our description goes beyond the treatment reported in e.g. Refs. \onlinecite{reittu1997,wortmann2001}, since we also include effects from the local spin-spin interaction between de-localized spin built up by the tunneling electrons, and the localized spin of the adatom. These spin-spin interactions are included already in the tunneling matrix element, c.f. Eq. (\ref{eq-me}), and describe that the tunneling electrons of different spins are subject to different tunneling barriers, i.e. spin-dependent tunneling barriers.

The theory presented here for obtaining quantitative spin-resolved information in a SP-STM experiment provides an alternative to the method used in Ref. \onlinecite{meier2008}, where a mean-field model based on the Weiss molecular field\cite{mohn2003} was used. Future work, primarily of experimental nature, will judge which approach is the most reliable one.

\acknowledgments
J.F. and O.E. thank the Swedish Research Council (VR) and the Royal Swedish Academy of Sciences (KVA), and A.V.B. thanks US DOE, for financial support. J.F. thanks LANL for hospitality during his visit in 2008. Special thanks to A. Bergman, S. Bl\"ugel, L. Nordstr\"om, B. Gy\"orffy, and W. Wulfhekel for valuable discussions, and to M. Bode and R. Wiesendanger for giving critical comments on the manuscript.
\\

\emph{Note added} | after the submission of the present paper the authors became aware of Ref. \onlinecite{lorente2009}.

\end{document}